\documentclass[twocolumn,showpacs,preprintnumbers,prc,superscriptaddress]{revtex4}
\usepackage{epsfig}
\usepackage{graphicx}
\usepackage{epstopdf}
\usepackage{amsmath,amssymb,amsfonts}
\usepackage{array}
\usepackage{url}
\usepackage{hyperref}
\usepackage{csquotes}
\usepackage{xspace}
\usepackage{textcomp}
\usepackage{ulem}
\usepackage[usenames,dvipsnames]{color}
\usepackage{setspace}

\newcommand{\sqsn}{\mbox{$\sqrt{s_{_{NN}}}$}\xspace}
\newcommand{\bef}{\begin{figure}}
\newcommand{\eef}{\end{figure}}
\newcommand{\bc}{\begin{center}}
\newcommand{\ec}{\end{center}}

\newcommand{\auau}{\mbox{Au$+$Au}\xspace}

\begin{document}
%\doublespacing
\title{ Study of diffusion coefficients of identified particles at energies
	available at BNL Relativistic Heavy Ion Collider }
\author{Vivek Kumar Singh}
\email{vkr.singh@vecc.gov.in}
\affiliation{Variable Energy Cyclotron Centre, HBNI, 1/AF Bidhannagar, 
Kolkata 700 064, India}
\author{Dipak Kumar Mishra}
\email{dkmishra@barc.gov.in}
\affiliation{Nuclear Physics Division, Bhabha Atomic Research Center, 
Mumbai 400085, India}
\author{Zubayer Ahammed}
\email{za@vecc.gov.in }
\affiliation{Variable Energy Cyclotron Centre, HBNI, 1/AF Bidhannagar, 
Kolkata 700 064, India}

\begin{abstract}
Using event-by-event fluctuations, we study the diffusion parameters of net-charge, net-pion, net-kaon, and net-proton in the heavy-ion jet interaction generator (HIJING), and ultra-relativistic quantum molecular dynamics (UrQMD) models at different collision energies \sqsn available at BNL Relativistic Heavy Ion Collider (RHIC). The diffusion parameter ($\sigma$) of net-charge and identified particles are estimated in rapidity space at various \sqsn. It is observed that, the $\sigma$ values are independent of collision energies but emphasises the particle-species dependence of diffusion coefficient in the QGP medium. The present work on particle-species dependence of diffusion coefficient provides a baseline for comparison with the experimental data.
% \pacs{25.75.Gz,12.38.Mh,21.65.Qr,25.75.-q,25.75.Nq}
\end{abstract}
\maketitle
\section{Introduction}
\label{intro}
Event-by-event fluctuations of conserved quantities such as net-charge, net-baryon number  etc. are widely studied in heavy ion collisions to understand the QCD phase transition~\cite{Shuryak:2000pd,Jeon:2000wg,Jeon:1999gr,Adams:2003st, Abelev:2008jg,ex3,Alt:2004ir,ex5,Abelev:2012pv,Asakawa:2000wh,Aziz:2004qu}. The variation of collision centrality, total energy deposition, variation in number of participant nucleons and baryon stopping in each and every collision causes these fluctuations. The strength of the measured fluctuations depends upon the survival probability in the QGP medium. However the fluctuations created in the initial state in heavy-ion collisions may survive during the hadronization process as the fireball expands very quickly~\cite{Asakawa:2000wh}. The strength of such fluctuations differs from their equilibrium hadron gas state towards initial values, typical for QGP~\cite{Asakawa:2000wh,Jeon:2000wg} if the relaxation time happens to be shorter than the lifetime of the hadronic stage of the collisions. The relaxation time scales of the fluctuations of different lengths or range in rapidity spaces are different. Since relaxation can only proceed via diffusion of the charge, the long-range fluctuations relax slower. The relaxation time grows as a square of the measured rapidity range~\cite{Shuryak:2000pd}. Thus the fluctuations of the total charge in a wider rapidity window relax slower. The minimal rapidity window we can consider must be much larger than the mean rapidity change of a charged particle in a collision, $\delta y_{coll}$. It is observed that, the typical $\delta y_{coll}$ for the baryon and electric charge is of the order 0.2 and 0.8, respectively. A large acceptance detector such as ALICE, STAR can be considered as the idealized limit of rapidity windows much broader than $\delta y_{coll}$. This can certainly be a reasonable acceptance to study the diffusion of the net-charge and identified particles.

One of the primary goals of the Beam Energy Scan (BES) program at RHIC is to explore the QCD phase diagram and transport properties of nuclear matter at different collision energy (indirectly varying temperature($T$) and net-baryon ($\mu_B$) density). The baryon chemical potential can reach upto $\mu_B\sim$ 400 MeV at lower collision energies available at RHIC. A strong gradients in the chemical potential of conserved charges are expected at such lower energies. Hence, lower energy beam scan at RHIC can be useful to explore the properties of charge particle diffusion of nuclear matter which were out of reach in high energy collisions.

\begin{figure*}[ht]
\bc
\includegraphics[width=1.0\textwidth]{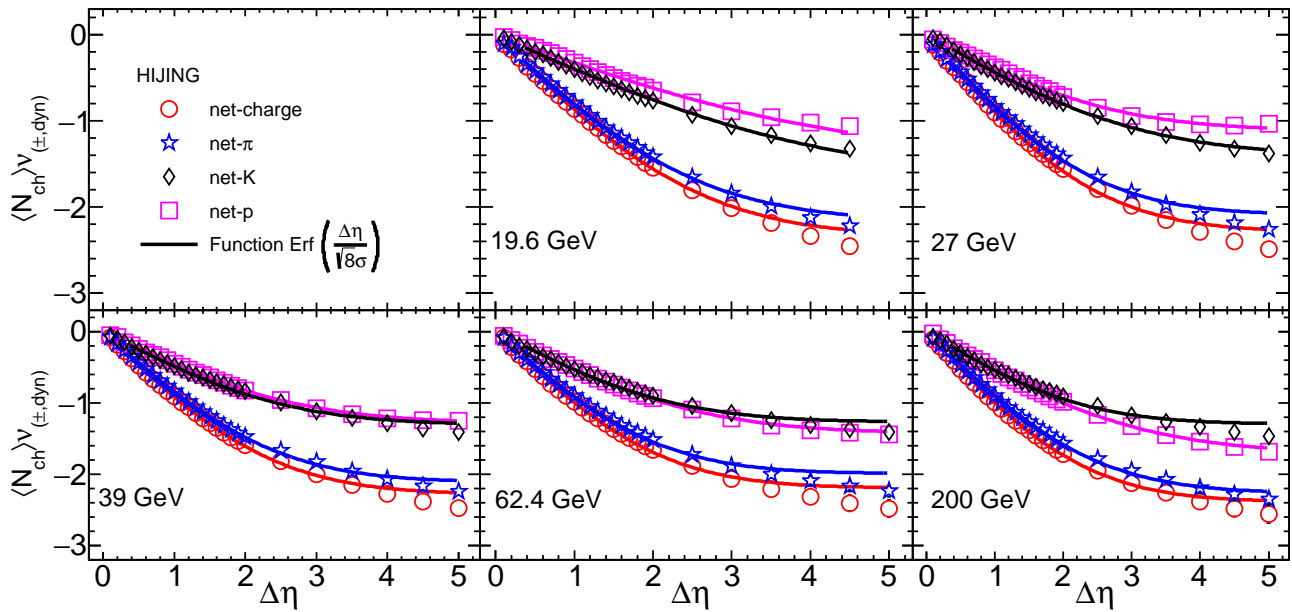}
\caption{The $\langle N_{ch}\rangle \nu_{(\pm,\mathrm{dyn})}$ for net-charge (red circle), net-pion (blue star), net-kaon (black diamond), and net-proton (magenta square) as a function of $\Delta\eta$ window for (0-5\%) centrality in \auau collisions at different \sqsn in HIJING model. The simulated data points are fitted with the $Erf(\Delta\eta/\sqrt{8}\sigma)$ from $\Delta\eta$ = 0.35 to 4.5 at 19.6 GeV and $\Delta\eta$ = 0.35 to  5.0 at 27--200 GeV respectively. The fitted curves are shown in solid lines. The statistical errors are within symbol size.}
\label{fig:NudynVsdeta@allE}
\ec
\end{figure*}

\begin{figure*}[ht]
\bc
\includegraphics[width=1.0\textwidth]{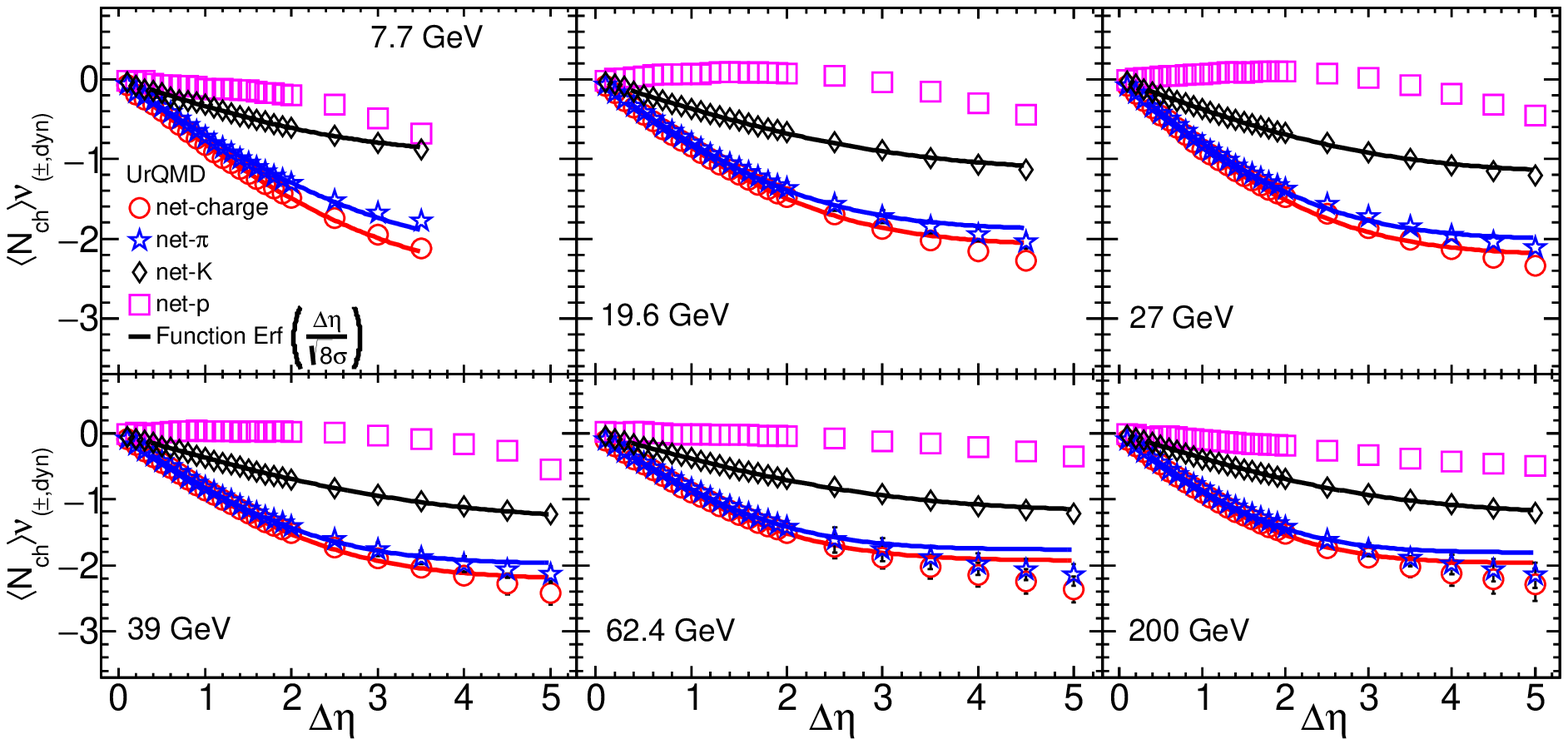}
\caption{The $\langle N_{ch}\rangle \nu_{(\pm,\mathrm{dyn})}$ for net-charge (red circle), net-pion (blue star), net-kaon (black diamond), and net-proton (magenta square) as a function of $\Delta\eta$ window for (0-5\%) centrality in \auau collisions at different \sqsn in UrQMD model. The simulated data points are fitted with the $Erf(\Delta\eta/\sqrt{8}\sigma)$ from $\Delta\eta$ = 0.35 to 3.5 at 7.7 GeV, $\Delta\eta$ = 0.35 to 4.5 at 19.6 GeV, and $\Delta\eta$ = 0.35 to  5.0 at 27--200 GeV respectively. The fitted curves are shown in solid lines. The statistical errors are within symbol size.}
\label{fig:NudynVsdeta@allEUrQMD}
\ec
\end{figure*}

The conservation laws limit the dissipation of the fluctuations which suffer after the hadronization has occurred. It is observed that these fluctuations may also get diluted in the expanding medium due to the diffusion of particles in rapidity space~\cite{Shuryak:2000pd,Aziz:2004qu,Abelev:2012pv}. The hadronic diffusion from time of hadronization $\tau_0$ to a freeze-out time $\tau_f$ can dissipate these fluctuations.  It is argued, one can observe the reduction of the fluctuation in QGP phase only if the fluctuations are measured over a large rapidity range~\cite{Shuryak:2000pd}. The QGP suppression of the charge fluctuations is not observed in the experimental data, while the suppression of charge fluctuations observed in experimental data is consistent with the diffusion estimates. Both the suppressions are crucially different from each other. While the QGP suppression is the history effect, the critical fluctuations are the equilibrium fluctuations pertaining to the freeze-out point, and the diffusion is necessary to establish them~\cite{Hatta:2003wn}.

Earlier efforts were made to estimate the fluctuation strength and diffusion parameter, $\sigma$ using net-charge of all-inclusive particles from a transport model and HRG model~\cite{Mishra:2017bdq}. Recently we have estimated net-charge fluctuation strength of different identified particles~\cite{Singh:2019skh}. However, the contribution of different identified particles to dilution of measured fluctuation strength and diffusion parameters may be different. It would be interesting to study the diffusion coefficient of different conserved charges in heavy-ion collisions. In Ref.~\cite{Greif:2017byw}, all diffusion coefficient matrix for baryon, electric charge and strangeness has been calculated using kinetic theory. The diffusion coefficient has also been calculated using net-charge fluctuations in heavy-ion data at LHC~\cite{Abelev:2012pv}. In the present study, we have considered the fluctuations in the production of all charged particles, as well as specific cases of identified pions, kaons, and protons and their antiparticle fluctuations. The former amounts to a measurement of net electrical charge fluctuations, whereas the latter corresponds to measurements of net-strangeness and net-baryon number fluctuations.

In this paper, we have calculated the diffusion coefficient parameter of identified charged particles, mainly for net-pion, net-kaon, and net-proton using a transport model UrQMD, and HIJING model which will serve as the baseline for experimental measurement. The paper is organized as follows. In the following section, we describe the formalism used to calculate the fluctuation strength. In Sec.~\ref{sec:diff models} we discuss HIJING and UrQMD model briefly. In Sec.~\ref{sec:results}, we discuss our estimated results on diffusion coefficients for identified particles. We finally summarise our findings in Sec.~\ref{sec:summary}.

\section{Measures}
\label{sec:Formalism}
The collisional volume can not be directly measured in heavy ion experiments, therefore the ratios of number of positive ($N_{+}$) and number of negative ($N_{-}$) charged particles normalized by total number of charged particles under consideration for a fixed centrality class of events is used to measure fluctuation strength. This quantity is usually known as $D$ measure~\cite{Jeon:2003gk} and defined as
\begin{eqnarray}
D =  \langle N_{\mathrm{ch}}\rangle\langle \delta R^2\rangle 
&=& \frac{4}{\langle N_{\mathrm{ch}}\rangle}\langle \delta N_+^2 + 
\delta N_-^2 - 2\delta N_+
\delta
N_-\rangle \nonumber \\
&\approx& \frac{4\langle \delta Q^2\rangle}{\langle N_{\mathrm{ch}}\rangle}
\label{eq:Dmea}
\end{eqnarray}
where $R (=N_+/N_-)$ is ratio of number of positive particles to number of negative particles. $Q = N_+ - N_-$ being the difference between number of positive and negative particles and $N_{\mathrm{ch}} = N_+ + N_-$ represents the total number of charged particles measured in an event. The $\langle \delta Q^2 \rangle$ is the variance of the net charge $Q$, which is proportional to the net charge fluctuation in the system. The value of $D$ is predicted to be approximately a factor of four times lower in the QGP phase as compared to the hadron gas phase~\cite{Jeon:2003gk}. However, the $D$-measure has been found to be dependent on detection efficiency. Hence, we use another variable, $\nu_{(\pm,\mathrm{dyn})}$ to measure the fluctuation strength. The $\nu_{(\pm,\mathrm{dyn})}$ is found to be robust and independent of detection efficiency. It is defined as
\begin{equation}
\nu_{(\pm,\mathrm{dyn})} = \frac{\langle N_+(N_+ - 1)\rangle}{\langle
	N_+\rangle^2} + \frac{\langle N_-(N_- - 1)\rangle}{\langle N_-\rangle^2} -
2\frac{\langle N_-N_+\rangle}{\langle N_-\rangle \langle N_+ \rangle},
\label{eq:nudyn}
\end{equation}

where $N_{+}$ is the number of positive particles and $N_{-}$ is the number of negative particles measured within the experimental acceptance. The measure of the relative correlation strength of (\enquote{$++$,} \enquote{$--$,} and \enquote{$+-$}) charged particle pairs is represented as $\nu_{(\pm,\mathrm{dyn})}$. The relation between $D$ and $\nu_{(\pm,\mathrm{dyn})}$ is given as~\cite{Jeon:2003gk}

\begin{equation}
\langle N_{\mathrm{ch}}\rangle \nu_{(\pm,\mathrm{dyn})}\approx D - 4
\end{equation}
The values of  $\nu_{(\pm,\mathrm{dyn})}$ need to be corrected for  global charge conservation and finite net charge effect~\cite{Pruneau:2002yf,Bleicher:2000ek}. However, in the present study we have refrained from from applying these corrections to our estimated $\nu_{(\pm,\mathrm{dyn})}$ values.

\section{Estimation of $\nu_{dyn}$ in HIJING and UrQMD models}
\label{sec:diff models}

In this section, we briefly discuss the two different heavy-ion models such as  
HIJING and UrQMD event generators used in this paper. The HIJING (V.1.37) and UrQMD (V.1.30) are used for the estimation of $\nu_{(\pm,\mathrm{dyn})}$ observable. Both HIJING and UrQMD models are the Monte Carlo event generators extensively used to explain the experimental data from high energy nucleon-nucleon and nucleus-nucleus collisions. These models provide a baseline to compare with the experimental data.  

HIJING is a perturbative QCD (pQCD) model based on the assumption of independent production of multiple minijets with initial and final state radiation in the collisions~\cite{Wang:1991hta}. The produced mini-jets are then transformed into string fragments and subsequently, fragments into hadrons. The model also incorporates the multiple jet processes and the nuclear effects such as multiple scattering, parton shadowing and jet quenching. PYTHIA is used to generate the kinetic variables for each hard scattering and the associated radiations, and JETSET is used for string fragmentation. The cross-sections for hard parton is calculated using leading order in order to account for higher order corrections. The soft beam jets are modeled using the diquark-quark strings with gluon kinks induced by soft gluon. The HIJING model considers the nucleus-nucleus collisions as a superposition of nucleon-nucleon collisions.

UrQMD is a classical transport model using free cross-sections for two-body interactions, which can be applied to study nucleon-nucleon, nucleon-nucleus and nucleus-nucleus interactions at relativistic energies~\cite{Bleicher:1999xi}. This model considers the transport based on covariant propagation of color strings, constituent quarks and diquarks with mesonic and baryonic degrees of freedom. Conservation of baryon number, electric charge, and strangeness number are preserved in the model. The phenomenology of hadronic interactions at low and intermediate energies (\sqsn $<$ 5GeV) describes in terms of the interactions between known hadrons and resonances. At higher energies, the multiple productions of particles describes in terms of the excitation of color strings and their subsequent fragmentation into hadrons~\cite{Bleicher:1999xi}. The model also considers the resonance decays, multiple scattering between hadrons during the evolution including baryon stopping phenomena, which is one of the important feature of heavy-ion collisions especially at lower collision energies~\cite{Bleicher:1999xi}. The UrQMD model has been applied successfully to study the hadron yields~\cite{Bass:1997xw,Soff:1999et}, event-by-event fluctuations and particle correlations~\cite{Bleicher:1998wu,Xu:2016qjd,Zhou:2017jfk,Netrakanti:2014mta, Westfall:2014fwa,He:2017zpg,Bleicher:2000ek,Li:2007yd}. A phase transition to a quark-gluon phase is not explicitly incorporated into the model dynamics.

It is to be mentioned that the measured values of fluctuation strength ($\nu_{dyn}$) depend on the width of the acceptance window, on the primordial mechanisms leading to $+ ve$ and $- ve$ particle production, radial transport (flow), diffusion, etc. Such effects are not explicitly taken into account in the both HIJING and UrQMD models.

\section{Results and discussion}
\label{sec:results}
The measured fluctuations may get diluted in the expanding medium due to the diffusion of the charged hadrons in the rapidity space~\cite{Aziz:2004qu}. The dissipation of fluctuations occurs during the evolution of the system from hadronization to their kinetic freeze-out. Hence, the experimental measurements of not only the magnitudes of fluctuation quantities at a fixed $\Delta\eta$ but also their dependence on $\Delta\eta$ enable us to explore various aspect of the time evolution of the hot medium and the hadronization mechanism. It is proposed to study the fluctuations of identified particle species and estimate the rate of diffusion in different rapidity interval for the measured particles at various centre of mass energies (\sqsn) available at BNL (RHIC). Measurements of the net-kaon and net-proton fluctuations are of particular interest as they address 
respectively fluctuations of net-strangeness and net-baryon number, which might 
be more sensitive to the details of the collision process.

\begin{figure}[ht]
\bc
\includegraphics[width=0.47\textwidth]{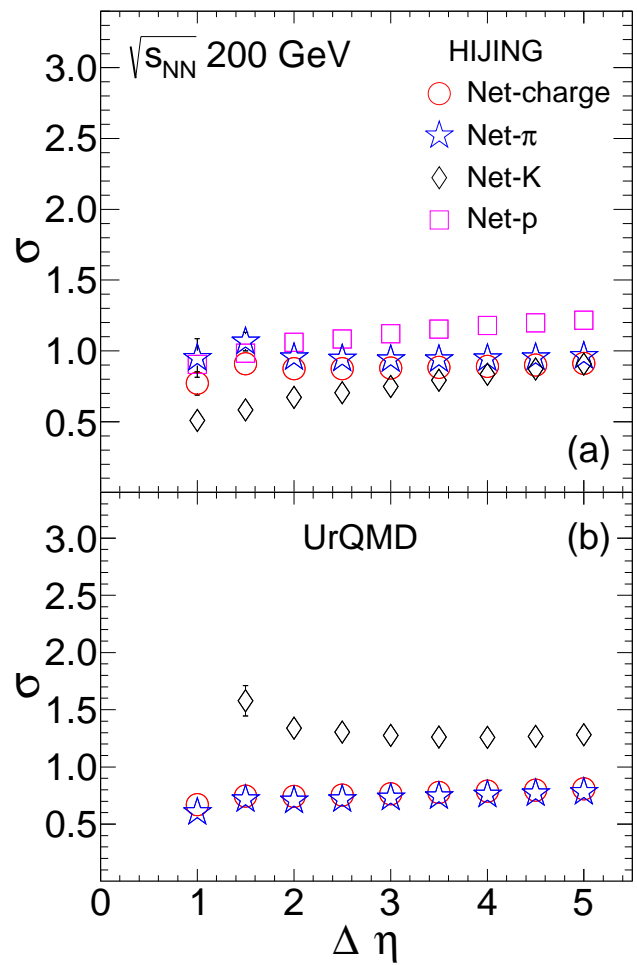}
\caption{Diffusion coefficient $\sigma$ as a function of $\Delta\eta$ window for 
net-charge (red circle), net-pion (blue star), net-kaon (black diamond), and 
net-proton (magenta square) are calculated using HIJING and UrQMD models for (0-5\%) centrality in \auau collisions. The statistical errors are within symbol size.}
\label{fig:sigmavseta200}
\ec
\end{figure}

\begin{figure}[ht]
\bc
\includegraphics[width=0.47\textwidth]{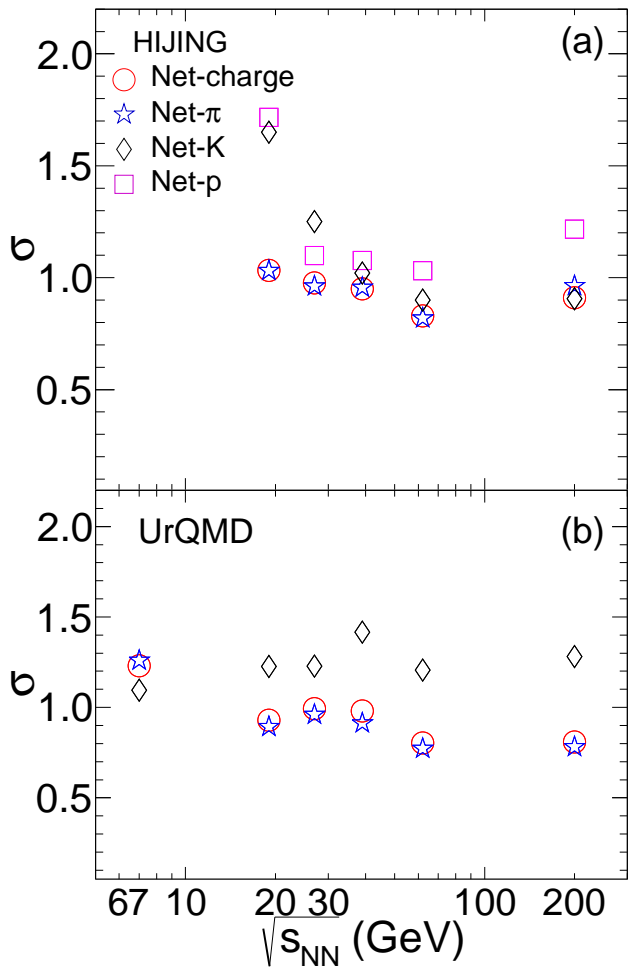}
\caption{The collision energy dependence of diffusion coefficient ($\sigma$) for
net-charge (red circle), net-pion (blue star), net-kaon (black diamond), and 
net-proton (magenta square) are calculated using HIJING and UrQMD models for (0-5\%) centrality in \auau collisions. The statistical errors are within symbol size.}
\label{fig:diff_sigma_ene}
\ec
\end{figure}

Figure~\ref{fig:NudynVsdeta@allE} and ~\ref{fig:NudynVsdeta@allEUrQMD} shows the $\langle N_{\mathrm{ch}} \rangle \nu_{(\pm,\mathrm{dyn})}$ as a function of $\Delta\eta$ intervals for (0-5\%) centrality in \auau collisions at different center of mass energies (\sqsn) using HIJING and UrQMD models, respectively. The dynamical fluctuations of the net-kaon and net-proton are larger than the dynamical fluctuations measured for net-pions and for inclusive non-identified net-charged particles. The net-proton dynamical fluctuations are somewhat larger than the net-kaon fluctuations. Following the Refs.~\cite{Abelev:2012pv,Aziz:2004qu}, the simulated data points are fitted with the error function, $\mathrm{Erf}(\Delta\eta/\sqrt{8}\sigma)$ representing the diffusion in rapidity space. The fitted functions are shown as solid lines in Figs.~\ref{fig:NudynVsdeta@allE} and \ref{fig:NudynVsdeta@allEUrQMD}. However, note that the width of the rapidity coverage is bound to include different physics phenomena at different beam energies. At 200 GeV, the beam rapidity is of the order 5.2 while at 7.7 GeV it is of the order of 2. Therefore, for \sqsn = 7.7, and 19.6 GeV, we consider $\Delta\eta$ upto 3.5 and 4.5 respectively. The data points are fitted within $\Delta \eta$ range, 0.35--3.5 at 7.7 GeV, 0.35--4.5 at 19.6 GeV, and 0.35--5.0 at other energies 27 to 200 GeV, for net-charge, net-pion, net-kaon and net-proton. The fit parameter, $\sigma$ in $\mathrm{Erf}(\Delta\eta/\sqrt{8}\sigma)$, characterizes the diffusion parameter at freeze-out that accounts for the broadening of the rapidity distributions due to interactions and particle production. We have calculated the diffusion coefficients at \sqsn = 7.7, 19.6, 27, 39, 62.4 and 200 GeV for all charge and the identified particles which represent the proxy for the conserved quantities (net-baryon, net-electric charge and net-strangeness) from both HIJING and UrQMD models. The slope of the fit function decreases with increasing particle mass.  

The $\Delta\eta$ dependence of $\langle N_{\mathrm{ch}} \rangle \nu_{(\pm,\mathrm{dyn})}$ for net-proton is qualitatively different in both HIJING and UrQMD models, whereas net-charge, net-pion and net-kaon have similar behavior in both the models. The fluctuation values from different models indicate that the fluctuations of charge is sensitive to the parton number embedded in the model. In the case of UrQMD model, the $\langle N_{\mathrm{ch}} \rangle \nu_{(\pm,\mathrm{dyn})}$ values of net-proton flattened at higher $\Delta\eta$ with increasing \sqsn. Hence, the $\langle N_{\mathrm{ch}} \rangle \nu_{(\pm,\mathrm{dyn})}$ as a function of $\Delta\eta$ for net-proton not able to fit with $\mathrm{Erf}(\Delta\eta/\sqrt{8}\sigma)$ function unlike pions and kaons. 

Figure~\ref{fig:sigmavseta200} shows the diffusion coefficient as a function of $\Delta\eta$ window for 
net-charge, net-pion, net-kaon, and net-proton for (0-5\%) centrality in \auau collisions at \sqsn = 200 GeV. 
The $\sigma$ values are obtained by fitting the $\langle N_{\mathrm{ch}} \rangle \nu_{(\pm,\mathrm{dyn})}$ up to different $\Delta\eta$ range with the error function. 
In both the HIJING and UrQMD models, the diffusion coefficient for net-charge and net-pion are 
independent of $\Delta\eta$ window and match with each other. This is due to the fact that the majority 
of the produced particles in a collision are pions itself. In HIJING model, the diffusion coefficients 
of net-kaon and net-proton show small $\Delta\eta$ dependence. The $\sigma$ values of net-proton are 
systematically above, whereas $\sigma$ values of net-kaon are systematically below the net-charge and 
net-pion values at all $\Delta\eta$ windows. In case of UrQMD model, the $\sigma$ values of 
net-kaon are higher than net-pion and net-charge. The difference in UrQMD with HIJING comes 
from the inclusion of multiple re-scattering  in the former. 
The HIJING model does not include the secondary interactions. The rescattering effect in UrQMD model has a substantial impact on all hardonic observables. In particular, the hadronic rescattering  leads to substantial de-correlation of the conserved charge distributions \cite{Urres}. As a result we find reduced diffusion coefficient in UrQMD with in comparison to HIJING model. 
Due to the qualitatively different nature of curvature of $\langle N_{\mathrm{ch}} \rangle \nu_{(\pm,\mathrm{dyn})}$ as a function $\Delta\eta$ for net-proton in UrQMD model, it was not possible to extract the $\sigma$ values.

The extracted values of diffusion coefficient of net charged and identified particles as a function of \sqsn are shown in Fig.~\ref{fig:diff_sigma_ene} from both HIJING and UrQMD models for (0-5\%) centrality in \auau collisions. The resulting values of $\sigma$ are obtained by fitting the $\langle N_{\mathrm{ch}} \rangle \nu_{(\pm,\mathrm{dyn})}$ values up to $\Delta\eta$ = 3.5 for 7.7GeV, 4.5 for 19.6GeV, and 5.0 for 27--200GeV respectively with the error function. The $\sigma$ values for net-charge and net-pion are close to each other at all the studied energies. The $\sigma$ of net-proton and net-kaon are closer to each other and systematically higher than net-pion at all energies in both the models. This indicates that the heavier particles are less diffused in the medium than the lighter particles. The number density decreases for the more massive species. As a result dilution  of the dynamical fluctuations reduces which goes inversely proportional to the multiplicity. Therefore, the heavier particles are less defused in comparison to the lighter one particles The strangeness conservation, and baryon number conservation also influence the size of the dynamical fluctuations for the net-kaon, and net-proton respectively. We observe that the diffusion coefficients are constant as a function of studied collision energy range \sqsn = 7.7 to 200 GeV. The results  remains of interest in so far as it provides references for the behaviour of diffusion parameter according to the models considered.

\section{Summary}
\label{sec:summary}
In summary, we have studied the fluctuations of net-charge, net-pion, net-kaon, and net-proton using the $\langle N_{\mathrm{ch}} \rangle \nu_{(\pm,\mathrm{dyn})}$ observable within the ambit of HIJING, and UrQMD models at different collision energies available at BNL (RHIC). The $\langle N_{\mathrm{ch}} \rangle \nu_{(\pm,\mathrm{dyn})}$ values are estimated up to higher $\Delta\eta$ window. In case of net-proton in the UrQMD model, the curvature of $\langle N_{\mathrm{ch}} \rangle \nu_{(\pm,\mathrm{dyn})}$ values as a function of $\Delta\eta$ shows different behavior. The diffusion coefficient ($\sigma$) has been estimated by fitting $\langle N_{\mathrm{ch}} \rangle \nu_{(\pm,\mathrm{dyn})}$ as a function of $\Delta\eta$ with the error function. It is found that, the $\sigma$ values are independent of collision energies. The $\sigma$ values of net-kaons and net-protons are systematically higher than net-pion at all the studied energies.  We also observe that the hadronic rescattering has significant effect on the correlation strength and the diffusion of the correlation in high multiplicity environment. The difference between HIJING and UrQMD is due to the fact that UrQMD, being a transport model, it includes multiple re-scattering before freeze out of the species which is absent in HIJING. This study set the baseline for the measurements at RHIC and the much awaited experimental data are needed in order to understand the particle production mechanisms.

\end{document}